\documentclass[%
twocolumn,
 amsmath,amssymb,
 aps,
]{revtex4-1}

\usepackage[utf8]{inputenc}
\usepackage{todonotes}
\usepackage{amsmath}
\usepackage{braket}
\usepackage{blindtext}  
\usepackage{cleveref}
\usepackage{color}

\usepackage{graphicx}% Include figure files
\usepackage{dcolumn}% Align table columns on decimal point
\usepackage{bm}% bold math
\begin{document}

\preprint{APS/123-QED}

\title{Emergent transport in a many-body open system driven by interacting quantum baths}
% Non-equilibrium* Phase Transitions induced by Strongly Coupled Interacting Bath}%in Strongly Coupled Spin Chains
% Bath Structure induced Nonequilibrium Phase Transitions
% Bath Structure/Interactions induced diffusion in strongly coupled spin chains

\author{Juris Reisons}
\affiliation{
Institute of Physics, Ecole Polytechnique F\'ed\'erale de Lausanne (EPFL), CH-1015 Lausanne, Switzerland}
\date{\today}

\author{Eduardo Mascarenhas}
\affiliation{
Institute of Physics, Ecole Polytechnique F\'ed\'erale de Lausanne (EPFL), CH-1015 Lausanne, Switzerland}
\date{\today}

\author{Vincenzo Savona}
\affiliation{
Institute of Physics, Ecole Polytechnique F\'ed\'erale de Lausanne (EPFL), CH-1015 Lausanne, Switzerland}
\date{\today}

\begin{abstract}
We analyze an open many-body system that is strongly coupled at its boundaries to interacting quantum baths. We show that the two-body interactions inside the baths induce emergent phenomena in the spin transport. The system and baths are modeled as independent spin chains resulting in a global non-homogeneous XXZ model. The evolution of the system-bath state is simulated using matrix-product-states methods. We present two phase transitions induced by bath interactions. For weak bath interactions we observe ballistic and insulating phases. However, for strong bath interactions a diffusive phase emerges with a distinct power-law decay of the time-dependent spin current $Q\propto t^{-\alpha}$. Furthermore, we investigate long-lasting current oscillations arising from the non-Markovian dynamics in the homogeneous case, and find a sharp change in their frequency scaling coinciding with the triple point of the phase diagram.

%We study spin transport in a boundary driven XXZ chain that is strongly coupled to interacting baths. Interacting baths are modeled as additional XXZ chains, and evolution of the system-bath state is simulated using matrix-product-states methods. We present two phase transitions induced by bath interactions: for bath interactions smaller than the spin hopping rate, we observe ballistic and insulating phases. By increasing bath interactions beyond spin hopping and system interactions respectively, both are replaced by a diffusive phase.

\end{abstract}

\maketitle

%\tableofcontents

\section{Introduction}

Non-equilibrium dynamics of quantum many-body systems have recently become the subject of considerable theoretical investigation. Of particular interest has been the question, foundational to quantum statistical mechanics, of equilibration and thermalization of many-body systems arising from unitary dynamics \cite{SpecialIssue,QuenchColloquim}. Largely responsible for this surge in interest are breakthroughs in experimental methods in the field of ultracold atoms, which make it possible to reproduce model Hamiltonians with great accuracy and investigate their unitary dynamics with unprecedented insulation from the environment \cite{Cold1,Cold2,Cold3,Cold4,Cold5,Cold6}.

From this context, the study of non-equilibrium phase transitions has emerged as a field of its own. These transitions differ significantly from equilibrium transitions in that they are not well understood as arising from thermal or quantum fluctuations\cite{QuantumPhase}, thus creating a need for new theoretical approaches \cite{Synergetics}. The study of transport in boundary-driven 1D systems provides a suitable paradigm to study these critical phenomena. The XXZ spin chain is an attractive choice for this purpose, both for its relative simplicity and ability to accurately describe real materials \cite{XXZ1,XXZ2,XXZ3}.

Transport in the XXZ model has been investigated under the assumption of Markovian coupling. At low bias near infinite temperature, where linear response theory is valid, diffusive and ballistic transport phases have been observed, with a transition at the Heisenberg point \cite{ZnidaricLinearResponse}. Investigations at high bias have instead revealed a ballistic and an insulating phase, separated by a subdiffusive Heisenberg point \cite{ProsenExactNess}.

The limitations of the Markovian approach are twofold. 
Firstly, the Markovian assumption is by definition valid only for weak coupling between system and bath. Secondly, in the case of weak coupling between system and bath but strong interactions within the system, a Markovian description is only available if one can obtain a full eigendecomposition of the system Hamiltonian, which may easily be beyond computational reach. Indeed, to derive the master equation by the book, all system operators in system-bath coupling should be expressed in the interaction picture, which results in expressing them in the basis of eigenoperators of the system Hamiltonian \cite{BreuerPetruccione}. If the couplings within the system are weak, the eigenoperators of the non-interacting system may be used as an ersatz, yielding a local-phenomenological master equation. Such an approach is however insufficient to model strong couplings, as has been recently shown by \cite{ChiralCurrents}.
Modeling both bath and system within a Hamiltonian formalism provides instead a way to investigate the regime of strong coupling and strong system interactions.

\begin{figure}
\includegraphics[scale=0.7]{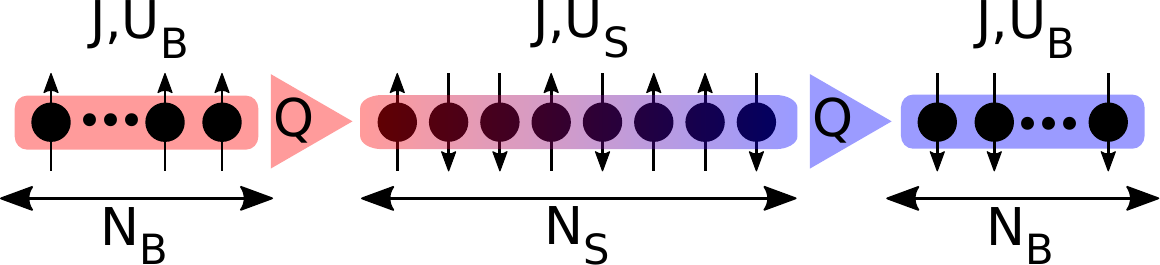}
\caption{Sketch of the chain. Spin current flows from left to right due to initial condition.}
\label{fig:sketch}
\end{figure}

Such a methodological shift has already proved fruitful, with several studies having investigated the dynamics of quenches arising from the junction of two spin chains. The junction of two XXZ chains has recently been shown to give rise to ballistic and diffusive transport phases \cite{SolvenianSingleJunction}. Motivated in part by the integrability of its dynamics, investigations of this setup have covered a large range of topics such as light cone velocities \cite{LightCone}, entanglement spreading \cite{EntanglementSpreading}, energy transport arising from joining chains of different temperatures \cite{Thermal1,Thermal2,Thermal3,Thermal4,Thermal5} and emerging hydrodynamics \cite{Hydrodynamics}. Two junction setups have also been studied : an XXZ chain coupled to two XX chains acting as magnetization reservoirs was found to behave similarly to the Markovian full-bias regime, with ballistic and insulating phases separated by a subdiffusive critical point. \cite{Giacommo}.

An additional opportunity opened by purely Hamiltonian evolution that has yet to be addressed is the possibility of investigating systems coupled to interacting baths. Indeed Markovian coupling requires the baths to be composed of non-interacting particles, and due to the prevalence of the Markovian paradigm in the field of open quantum systems the effects of interactions in the baths have been left mostly unexplored. %The introduction of interactions has the potential to give rise to novel transport phenomena, and since their modeling requires little additional effort if we resort to Hamiltonian evolution we are naturally motivated to investigate their effects.
In this work, we present evidence of critical behavior arising from bath interactions in a strongly coupled boundary-driven spin chain.

\begin{figure}
\includegraphics[scale=0.3]{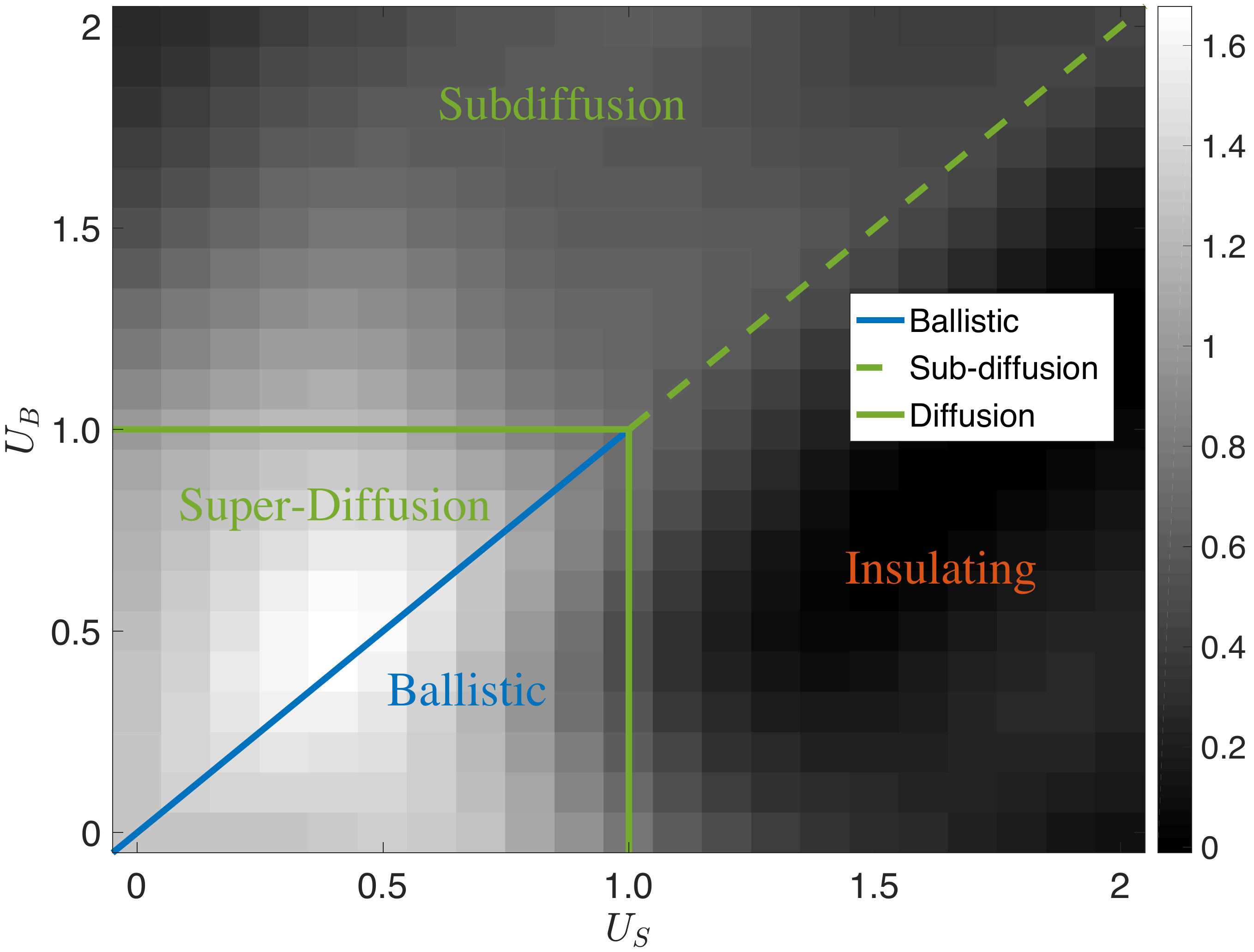}
\caption{Junction currents as function of $U_S$ and $U_B$ for system size $N_S = 20$. Currents are taken during the second transient regime $Q(\tau_{1\rightarrow2})$,$\tau_{1\rightarrow2} = 0.15N_B/J$. Superimposed, proposed phase boundaries. The transport of each phase boundary is indicated by the legend. The type of transport in each area-region delimited by the boundaries is indicated by the text written on top of the figure.}
\label{fig:Phase_diagram}
\end{figure}

\section{Model}

We study the dynamics of a tripartite XXZ chain, which is sketched in Fig. \ref{fig:sketch}. The first and third parts of the chain play the role of positive and negative leads of a magnetization battery, and will be referred to as the battery leads. The middle part will be referred to as the system. We call $N_B$ the length of the batteries and $N_S$ the length of the system. Sites $N_B$ and $N_S + N_B$ are situated at the interfaces of battery leads and system, and will be referred to as the junctions. Unless specified otherwise, $N_B = 1.5N_S$.

The Hamiltonian for the entire chain can be expressed in terms of Pauli matrices as
\begin{equation}
H = \sum_{i}^{2N_B + N_S - 1} J(X_{i}X_{i+1} + Y_{i}Y_{i+1})  + U_iZ_{i}Z_{i+1}
\end{equation}
\begin{equation*}
U_i =
\begin{cases}
    U_B ,& \text{if } i \leq N_B \text{ or } i \geq N_B+N_S \\
    U_S,              & \text{otherwise}
\end{cases}
\end{equation*}
with $J$ the spin hopping rate and $U_B$,$U_S$ the spin repulsions inside the battery leads and system respectively.

At the start of the simulation, we prepare the battery leads in the $\ket{\uparrow\uparrow...\uparrow\uparrow\uparrow}$ and $\ket{\downarrow\downarrow...\downarrow\downarrow\downarrow}$ states. The system is prepared in the ground state of its XXZ Hamiltonian $\ket{G}$. The initial state of the whole chain is thus $\ket{\Psi} = \ket{\uparrow\uparrow...\uparrow\uparrow\uparrow}\ket{G}\ket{\downarrow\downarrow...\downarrow\downarrow\downarrow}$.

The dynamics resulting from this initial state can be understood as the result of two local quenches occurring at the junctions. These quenches spawn excitations that propagate throughout the chain.

Our global Hamiltonian being non-homogeneous, it is not solvable by Bethe Ansatz techniques. We rely instead on DMRG methods, which have proven efficient at simulating local quenches. Simulation of the system is performed using time-dependent matrix product state techniques (tMPS). Time evolution is performed using second-order Trotter-Suzuki decomposition with time step $dt = 0.05/J$ and maximal bond dimension $D = 500$. 

The transport properties are studied by computing the spin currents $ Q_i = 2J\braket{X_iY_{i+1} - Y_iX_{i+1}}$, which appear in the continuity equation $\braket{\dot{Z_i}} = Q_{i-1} - Q_{i}$. Of particular interest are the current at the positive lead junction which we denote as $Q$ and the current in the middle of the system, $Q_m$. The time dependence of $Q$ reveals two distinct transient regimes. We note $\tau_1$ and $\tau_2$ the end of each transient regime, and $Q(\tau \geq \tau_2)$ the quasi-steady-state current.

\begin{figure*}
\includegraphics[scale=0.5]{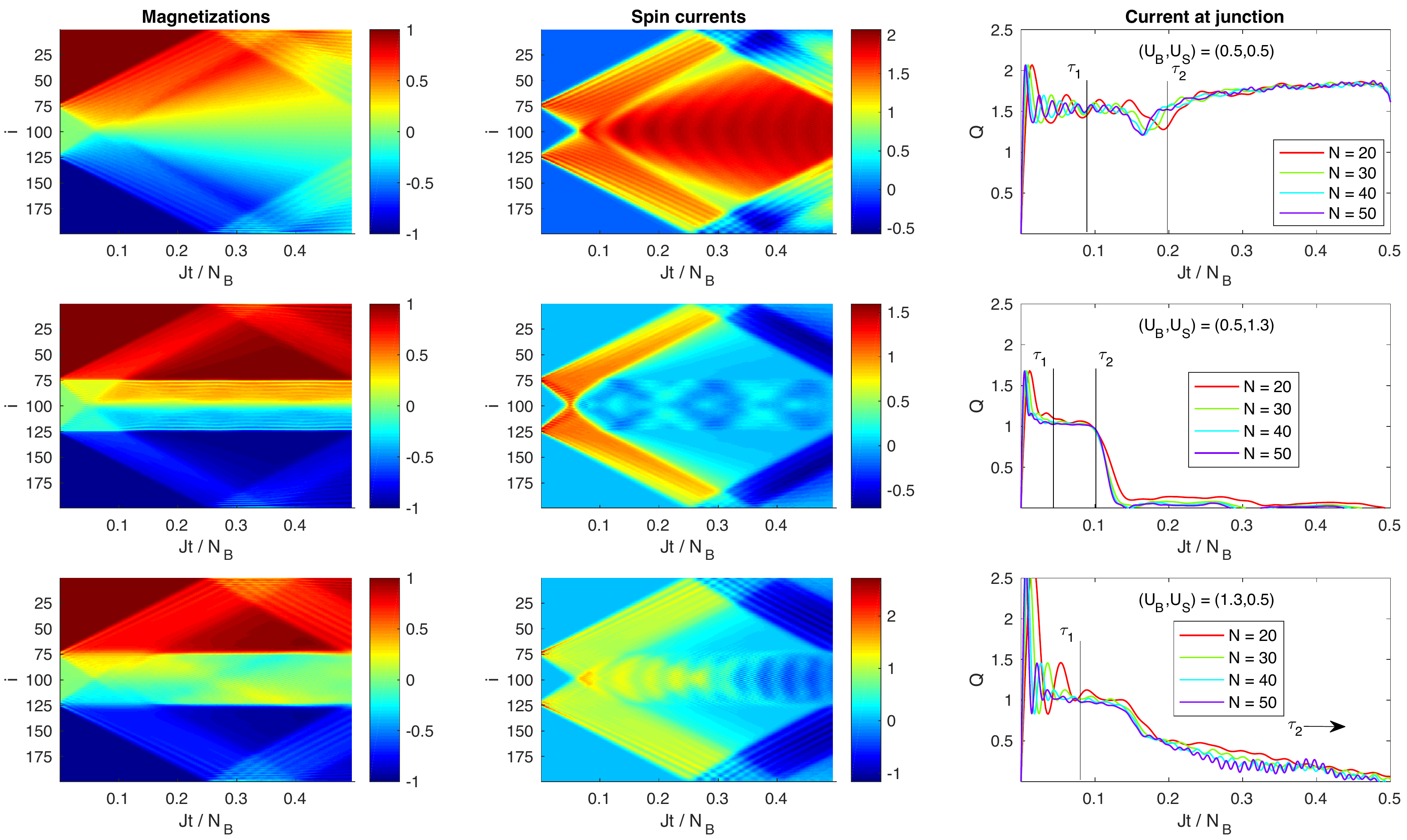}
\caption{Color plots of Magnetization $\braket{Z_i}$ and spin current $\braket{Q_i}$ as functions of time and space for $N_S = 50$, spin current at battery - system junction as function of time for various system sizes. Top row : $U_B,U_S = (0.5,0.5)$. Middle row, $(0.5,1.3)$. Bottom row, $(1.3,0.5)$.} 
\label{fig:AllPhases}
\end{figure*}

 In previous literature there have been two main strategies for characterising the type of transport. The spreading of local excitations~\cite{SolvenianSingleJunction} and the system size scaling of persistent currents~\cite{ZnidaricLinearResponse,ProsenExactNess}. We note that both approaches may be directly linked as discussed in~\cite{PhysRevLett.117.040601} and in principle only one of the above criteria should be enough to characterize the transport, however we have found in practice by performing both analyses that the time behaviour gives more consistent conclusions. 

Let us assume the current to scale with the system size as $Q\propto N^{-\gamma}$. We have that if $\gamma=0$ the system is a perfect ballistic conductor, $\gamma<1$ indicates super-diffusion, $\gamma=1$ diffusion and $\gamma>1$ super-diffusion. This also translates into the time behaviour of the current. In the spirit of spreading of inhomogeneities we consider the total magnetization transferred form one of the baths 
\begin{equation}\Delta Z(t)=\int_0^t Q(\tau)\propto t^{\delta} \end{equation}
such that $\delta=1$ indicates ballistic transport, $\delta>\frac{1}{2}$ super-diffusion, $\delta=\frac{1}{2}$ diffusion and $\delta<\frac{1}{2}$ sub-diffusion. Furthermore, if $Q(t)\propto t^{-\alpha}$ we may identify $\alpha=1-\delta$. A relation between $\gamma$ and $\alpha$ maybe expected, however we find no obvious functional form.

We point out that the phenomenological master equation driving in~\cite{ZnidaricLinearResponse,ProsenExactNess,PhysRevLett.117.040601} ensures persistent currents even outside the ballistic phase. In contrast, our simulations that explicitly model the bath do not guarantee that currents will persist in the infinite time limit. Therefore, even though we make an effort to relate the current work to the finite size scaling in~\cite{ProsenExactNess} we find the time behaviour of the current to be a more appropriate object for study.

\section{Conjectured Phase Diagram}

In Fig. \ref{fig:Phase_diagram} we present junction currents in the second transient regime as a function of $U_S$ and $U_B$ obtained for a system of size $N_S = 20$. At a glance, one sees a square area of high current defined by $max(U_B,U_S) < 1$. We show that this area exhibits ballistic transport at and below the line $U_S = U_B$ line and super-diffusive transport above the line. Outside the square, another separation can be seen along the $U_S = U_B$ line, with much greater current above it than below. This motivates us to distinguish two additional phases: a sub-diffusive phase above the line and an insulating phase below. We show that the current $Q(\tau)$ has power law time decay in the generalized-diffusive phases but exponential decay in the insulating phase. It should be noted that the anamalous-diffusive phases are a novel feature, contingent on the presence of interactions in the bath.
The above description of the phase diagram is specific regarding the type of diffusion found in each region. However, our focus here is not the precise determination of the anomalous diffusion exponents since these are also plagued by numerical and finite-size effects. Therefore, we note that in some cases we refer to all the diffusive-type phases simply as diffusive when it comes to differentiating them with respect to the ballistic and insulating phases.

Fig. \ref{fig:AllPhases} presents magnetization and current profiles characteristic of the three phases. A few general features of the dynamics can be noted. 
In all phases, one can see two light cones arising from the quenches at the junctions. This structure gives rise to two transient regimes of the junction current. The first regime lasts until the light cone from one junction crosses the system and hits the opposite junction. We refer to this time as $\tau_1$. Behavior in all phases is similar in this regime: current starts to flow from both leads into the system. The dynamics of this regime are those of a single battery-system junction.

It is instead the second transient regime and the quasi-steady-state that reveal the differences between the phases. In contrast to the first transient, their behavior is dictated by the interference of the two light cones. In the ballistic phase, the merging of the light cones gives rise to a finite value of the current and a smooth magnetization profile. In the insulating phase, we instead observe destructive interference causing a sharp drop of the current to 0. The magnetization profile displays staggered order in the system and a sharp step of the magnetization profile in the middle. In the diffusive phase, we observe instead a remarkably different evolution of the profile. The magnetization gradient in the system can actually be reversed, with $\braket{Z_i} < 0 $ close to the positive lead, and vice-versa at the negative lead. In addition, the net drop of the current to 0 is much slower with fast oscillations. These differences in the current time-dependence and magnetization profiles provide evidence that the diffusive phase is a novel phase induced by bath interactions. To complement these qualitative observations, we provide a finite-size scaling analysis of the ballistic-insulating and ballistic-diffusive transitions, as well as quantitative evidence for the distinct dynamical signatures at the insulating-diffusive transition.

%\subsection{Ballistic - Insulating transition}
\label{sub1}

\begin{figure}
\includegraphics[scale=0.2]{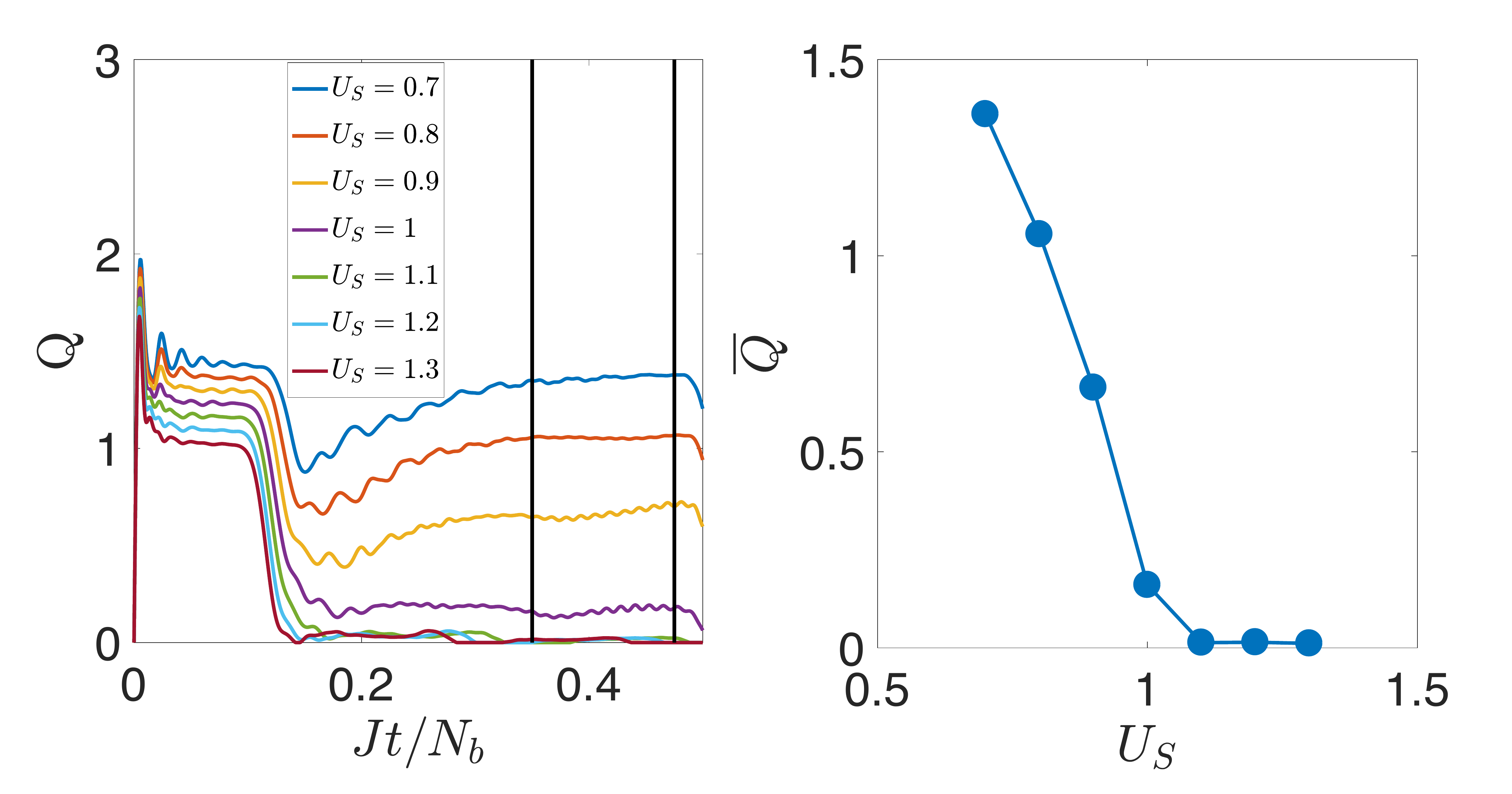}
\caption{(Left) The current $Q(t)$ as a function of time for a central system of $N=50$ with a bath interaction of $U_B=0.5J$. (Right) The long time average current $\overline{Q}$ as a function of the system interaction. The time interval for averaging is indicated in the left panel.} 
\label{Insulating_dynamics}
\end{figure}

\begin{figure}
\includegraphics[scale=0.17]{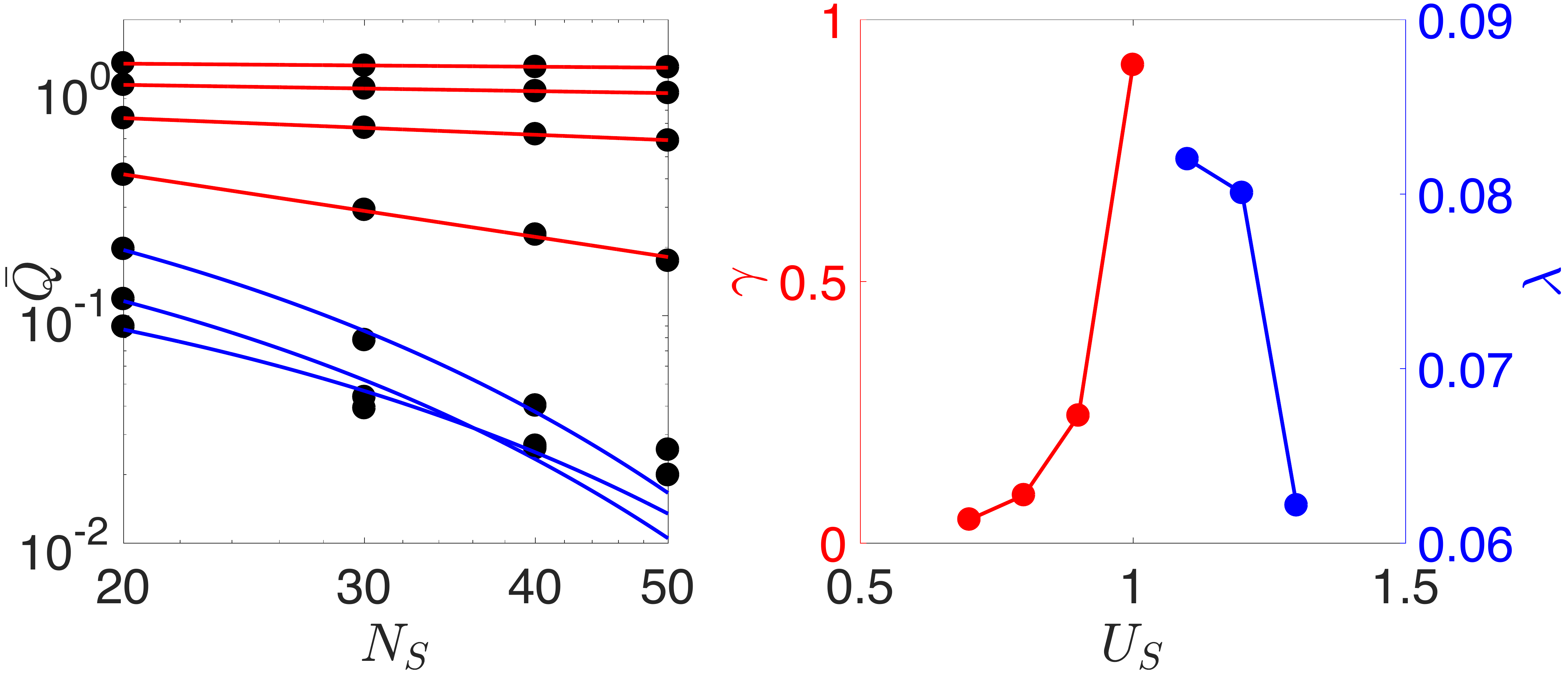}
\caption{Data points and associated power-law $\bar{Q}(N) = AN^{-\alpha}$ and exponential $\bar{Q}(N) = Be^{-\beta N}$ fit results for $U_B = 0.5$, $U_S \in [0.7,1.3]$. The time-averaging interval is $[\tau_2 = 0.15,T=0.45]N_B/J$} 
\label{fig:B_S_scaling}
\end{figure}

In FIg.~\ref{Insulating_dynamics} we address the ballistic-insulating transition and we show the time evolution of the current $Q$ that takes persistent non-vanishing values only for $U_S\le1$. The dynamical behaviour given a very sharp indicator of the transition.

Finite-size scaling of quasi-steady state current was investigated along the ballistic-insulating transition. The Non-Markovian dynamics induce oscillations of the current around its average even at long times. For this reason we fit the time-averaged current $\bar{Q} = \frac{1}{T-\tau_2}\int_{\tau_2}^{T} Q(t)dt$ with respect to system size . Fig. \ref{fig:B_S_scaling} presents the results of a power-law fit for the ballistic phase and exponential fit for the insulating phase.

For $U_S < 1$, the vanishing exponent is a clear indication of system-size independence and ballistic behavior. However, we recognize finite size effects give a small but non zero exponent especially when closer to the transition at $U_S=1$. At the transition, we observe approximately normal diffusion $\gamma \approx 0.9$. Above the transition point the values of the current are small and MPS truncation errors become relevant, especially for large system sizes. Our scaling data would suggest very weak diffusion, however due to the dynamical fast drop of the current in this regime our best interpretation is that an exponential scaling emerges: in this circumstance, we consider dynamical features to be better indicators than the scaling. This motivates our choice of reporting the exponential fits in Fig. \ref{fig:B_S_scaling}. All these findings are similar to what was found in the $U_B = 0$ case in \cite{Giacommo}, suggesting the bath interaction plays no meaningful role in this region of the phase diagram.

%\subsection{Ballistic - Diffusive transition}
\label{sub2}

\begin{figure}
\includegraphics[scale=0.2]{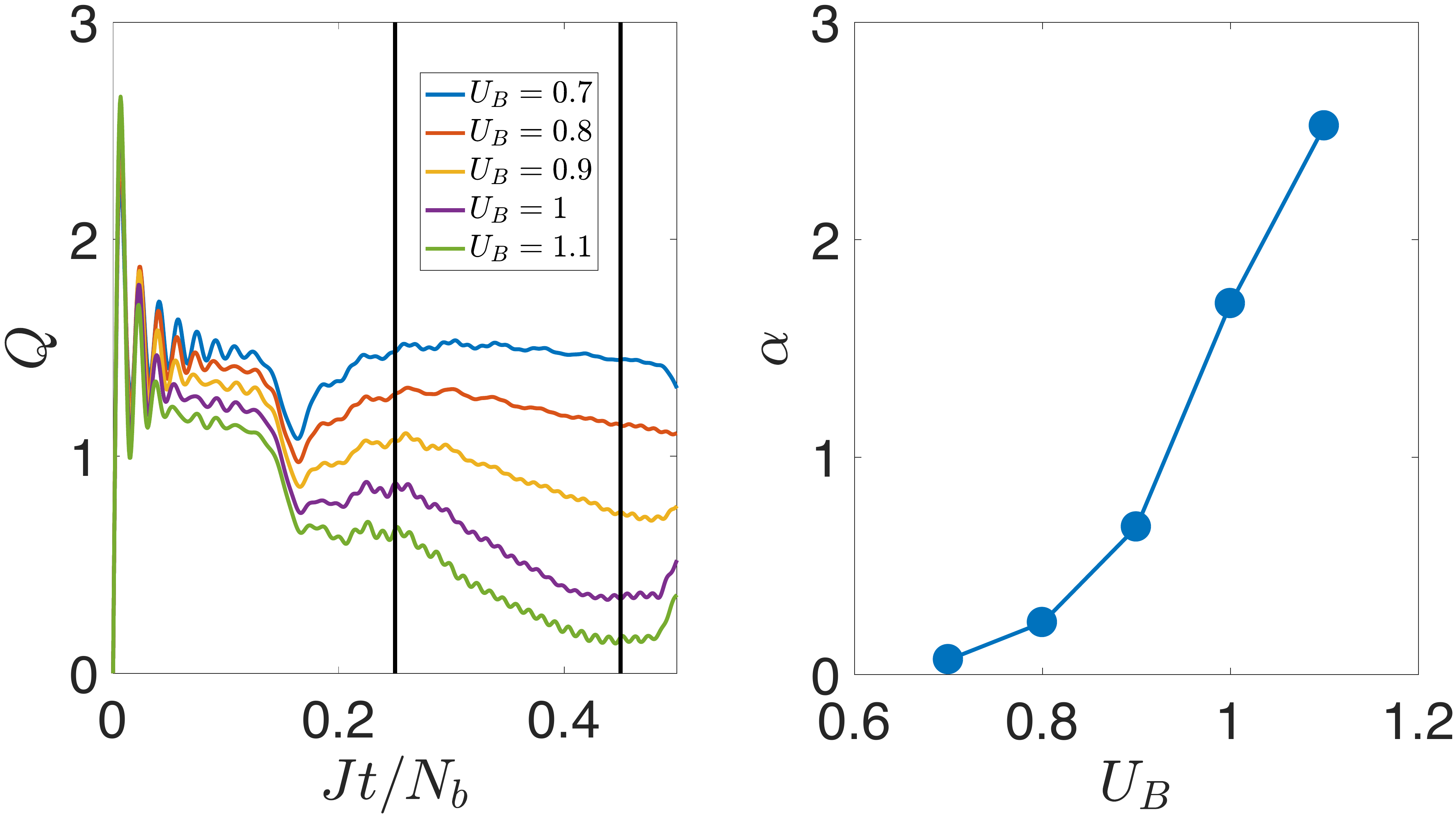}
\caption{(Left) The current $Q(t)$ as a function of time for a central system of $N=50$ with a system interaction of $U_S=0.5J$. (Right) The exponent of the power law $Q\propto t^{-\alpha}$ as a function of the system interaction. The time interval for fitting is indicated in the left panel.} 
\label{DiffusiveDynamics}
\end{figure}

\begin{figure}
\includegraphics[scale=0.17]{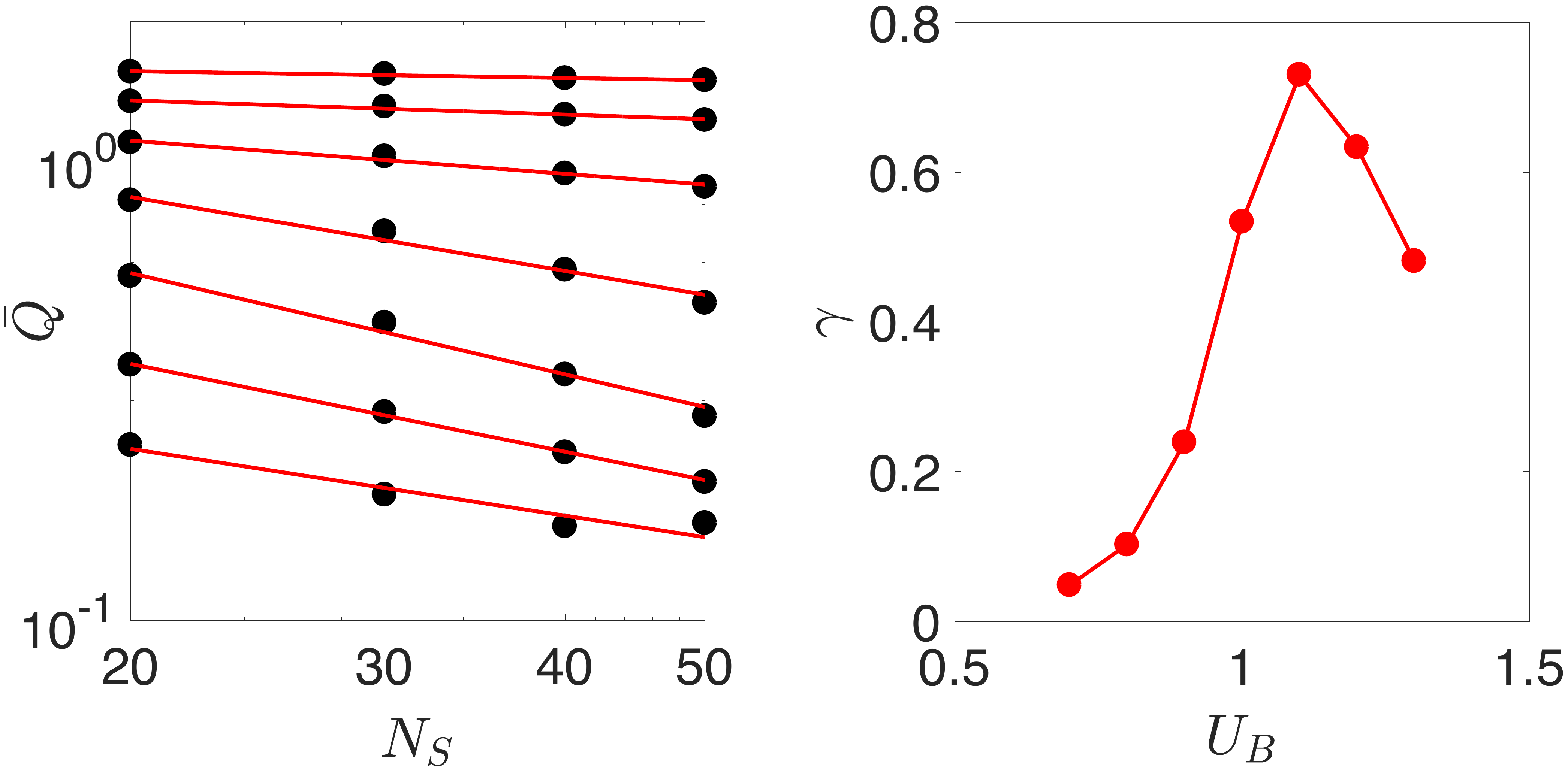}
\caption{Data points and associated power-law $\bar{Q}(N) = AN^{-\alpha}$ fit results for $U_B \in [0.7,1.3]$, $U_S =0.5$. The time-averaging interval is $[\tau_2 = 0.3,T=0.45]N_B/J$}
\label{fig:B_B_scaling}
\end{figure}

The same procedure was performed for the ballistic-diffusive transition. 
In Fig.~\ref{DiffusiveDynamics} we show the time dynamics across the ballistic-diffusive transition as we increase the bath interaction. Decaying power-laws emerge as we approach $U_B=1$. Our data suggests the diffusion point to be at $U_B\approx 0.85$, however due to finite size simulations and truncation errors are results do not allow us to draw the precise location of the diffusive point nor weather or not the weak power laws preceding it are just due to finite size effects. What we have certainly stablished is that for very small $U_B$ the system is a ballistic conductor and as we increase $U_B$ the system turns into diffusive and even sub-diffusing conductor.
Results are presented for the finite size scaling in Fig. \ref{fig:B_B_scaling}. We find weak system-size dependence for $U_B < 1$, which becomes stronger approaching $U_B \geq 1$. Here however, the finite size scaling seems to be a less meaningful analysis. Our analysis would suggest super-duffusive behaviour however, the fast power-laws in Fig.~\ref{DiffusiveDynamics} indicates sub-diffusion. Regardless of the precise exponents and transition point the fact the the bath interactions induce generic diffusive behaviour in evident.

%\subsection{Insulating - Diffusive transition}
\label{sub3}
\begin{figure}
\centering
\includegraphics[scale=0.17]{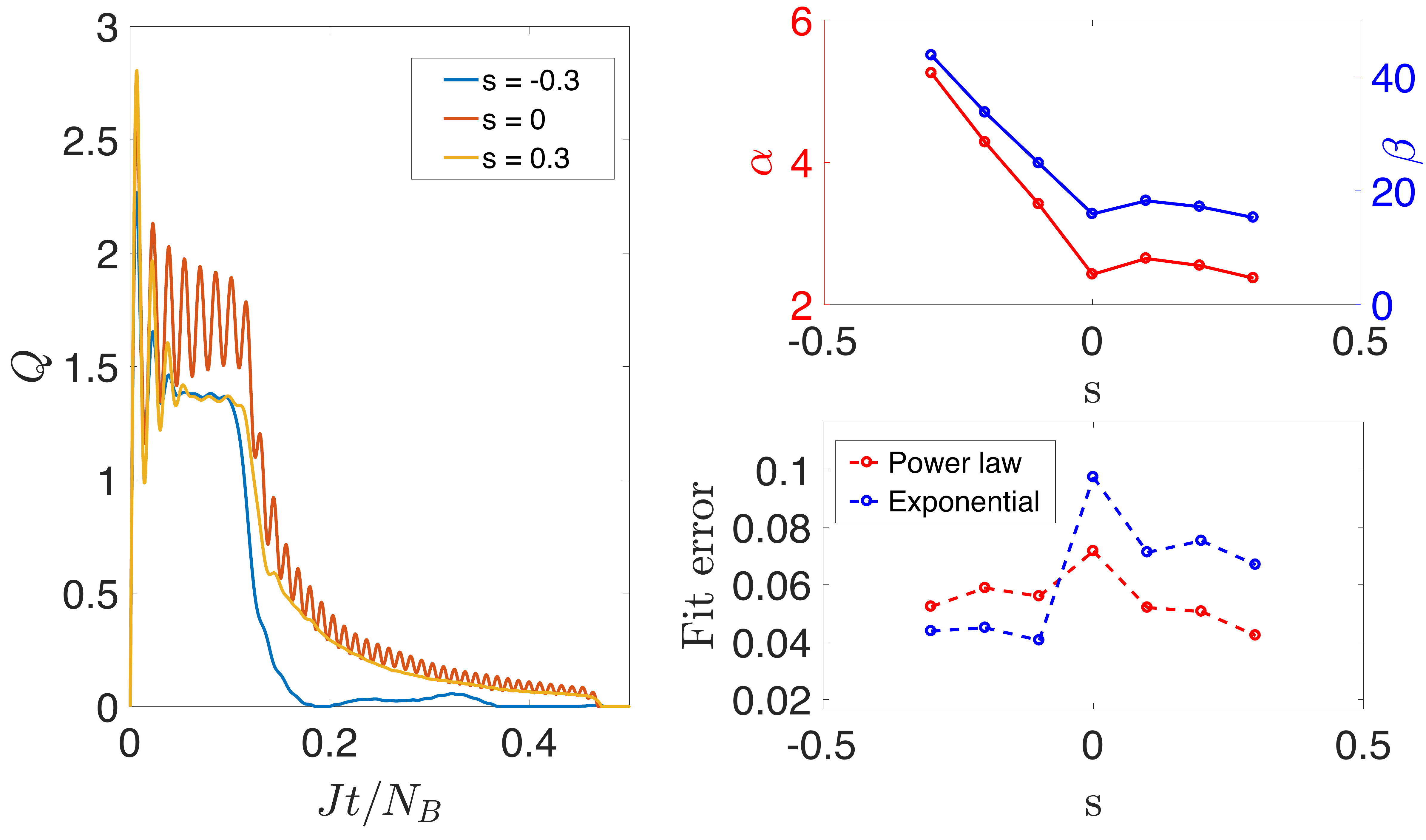}
\caption{On the left, junction current as function of time for point in insulating phase (s=-0.3), diffusive phase (s=0.3), and on the transition line (s = 0),$N_S = 50$. On the right, results of power law $Q(t) = At^{-\alpha}$ and exponential $Q(t) = B\exp^{-\beta t}$ fits on the $Jt/N_b \in [0.1,0.45]$ time interval for $s \in [-0.3,0.3]$.} 
\label{fig:time_fit}
\end{figure}

We now turn to quantifying the time-dependence difference between the diffusive and insulating phase.
We investigate data points on a line perpendicular to $U_S = U_B$, which we parametrize by $s$ as $\begin{pmatrix}U_B\\U_S\end{pmatrix} =\begin{pmatrix}1.5\\1.5\end{pmatrix} + s\frac{1}{\sqrt2} \begin{pmatrix}1\\-1\end{pmatrix}$.
The results are presented in Fig. \ref{fig:time_fit}. 

On the left, current as a function of time is drawn for a point in each phase and a point on the diagonal for $N_S = 50$. Inside the insulating phase ($s=-0.3$) we can see the dynamical signature of this regime which is the fast drop of the current towards zero. Exactly at the diagonal ($s=0$) we have the transition point in which we can see two distinct features. Persistent fast oscillations are the trait of the transition point. These oscillations carry, however, an envelope given by a time-algebraic decay $Q\propto t^{-\alpha}$ which is the signature of the novel diffusive phase that extends above the diagonal ($s=0.3$). We have fitted the time evolution of the current after the interference of the light cones both with an exponential and a power law.
The Power-law and exponential fit coefficients and errors are presented on the right. As one can see from the fit errors, $s <0 $ is better described by an exponential decay, while at the transition and beyond $s \geq 0$ the power-law is a better description. The two features are consistent with insulating and diffusive transport respectively, and therefore concur with the results of finite-size scaling. 

\begin{figure}
\includegraphics[scale=0.2]{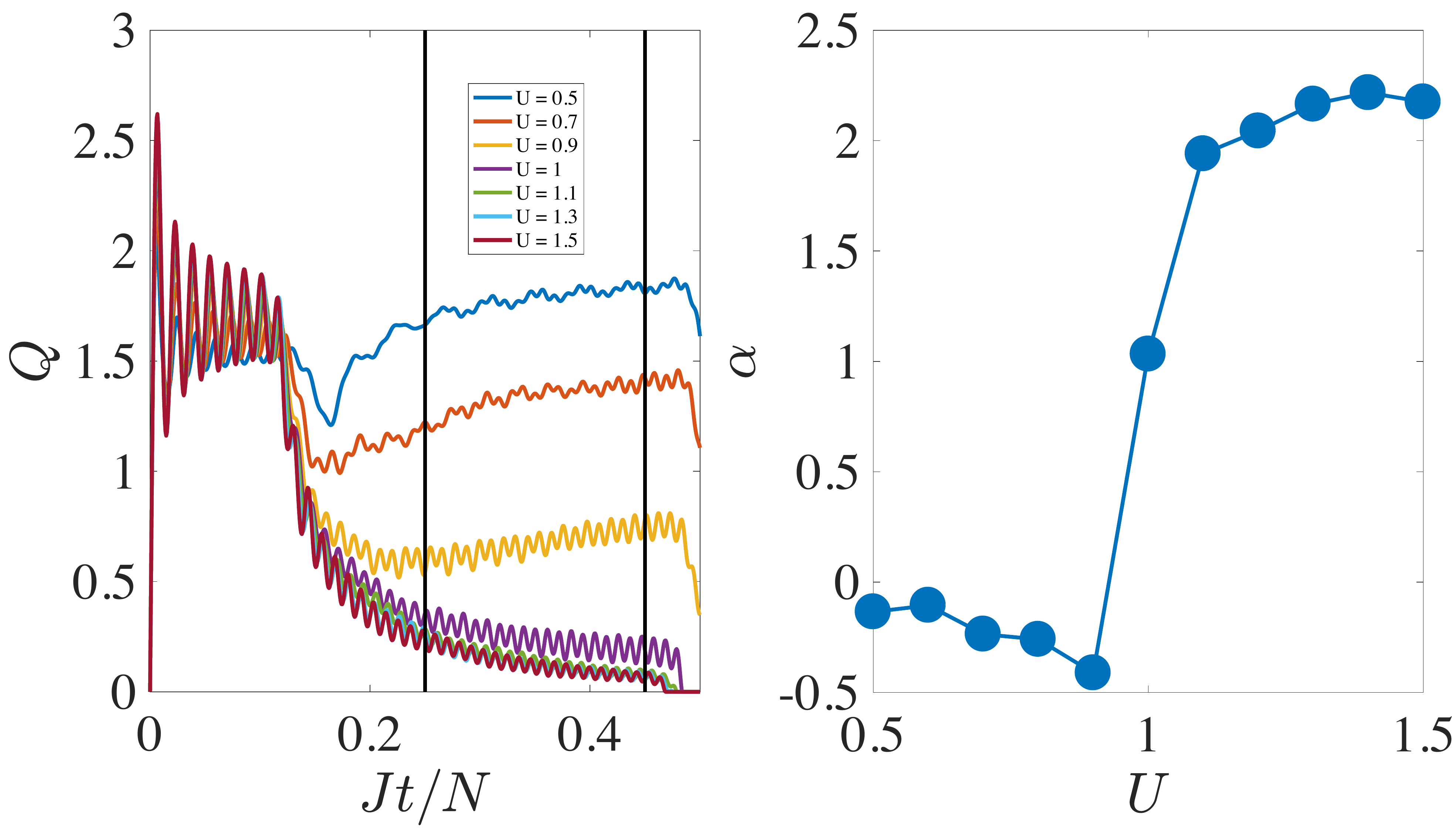}
\caption{(Left) The current $Q(t)$ as a function of time for a central system of $N=50$ with a system interaction of $U_S=U_B=U$. (Right) The exponent of the power law $Q\propto t^{-\alpha}$ as a function of the interaction. The time interval for fitting is indicated in the left panel.} 
\label{HomogenousDynamics}
\end{figure}

Finally, we compare our protocol to the one in~\cite{SolvenianSingleJunction} in the case in which bath and system interactions as the same resulting in a homogeneous hamiltonian with different inhomogeneous initial conditions. Our results in Fig.\ref{HomogenousDynamics} indicate ballistic transport below $U<1$ with a sharp transition to sub-diffusion while the results in~\cite{SolvenianSingleJunction} indicate normal diffusion. We not that our finding do not contradict ~\cite{SolvenianSingleJunction} since the initial conditions are markedly different.

\section{Oscillations and trapped quasi-particles}

As one can notice in Fig. \ref{fig:AllPhases}, oscillations of the current appear in the the system after the light cones collide. The domain where these oscillations occur is identical to the domain where magnetization is close to 0. Thus, these oscillations spatially expand in the ballistic phase, but remain localized inside the system in both it the diffusive and insulating phases.

\begin{figure}
\centering
\includegraphics[scale=0.17]{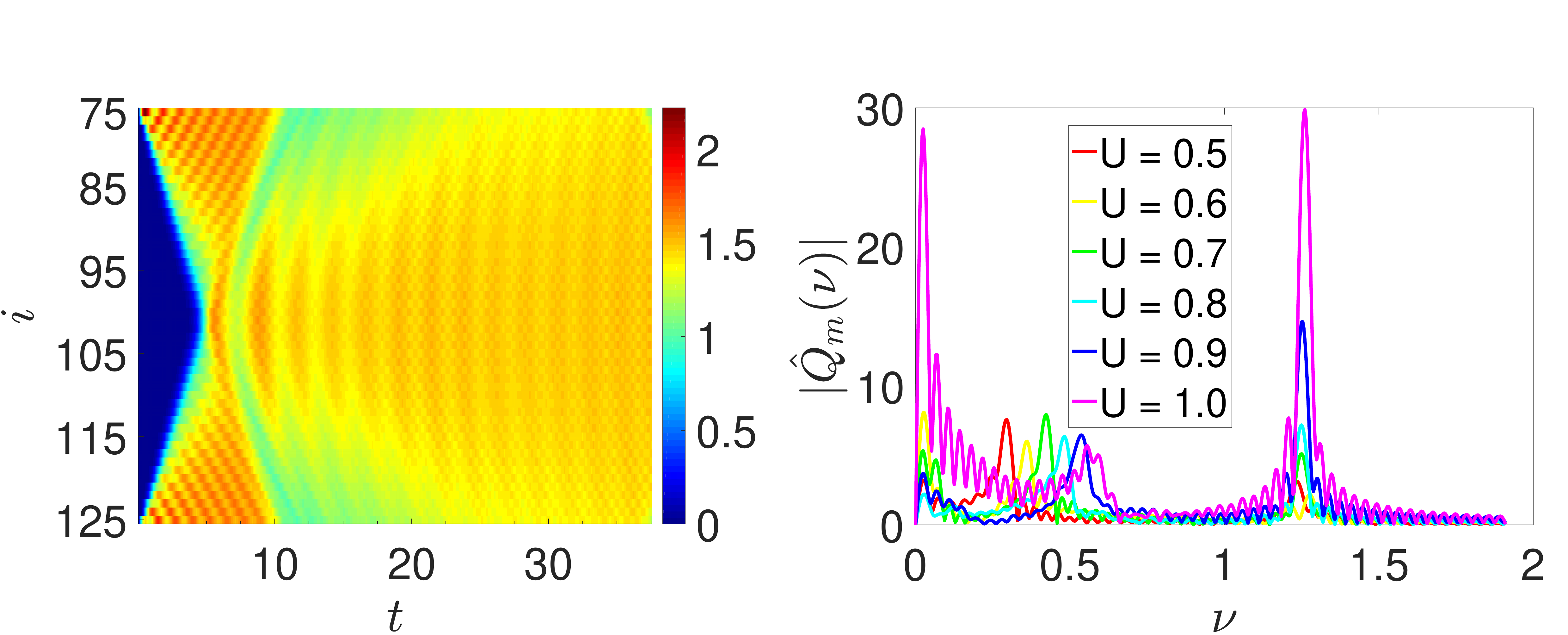}
\caption{Left : Current profile for $U = 0.7$. High and medium frequency oscillations are well visible. Right : $|\hat{\tilde{Q}}_m(\nu)|$ for points in ballistic phase.}
\label{Fourier_Ballistic}
\end{figure}

\begin{figure}
\centering
\includegraphics[scale=0.17]{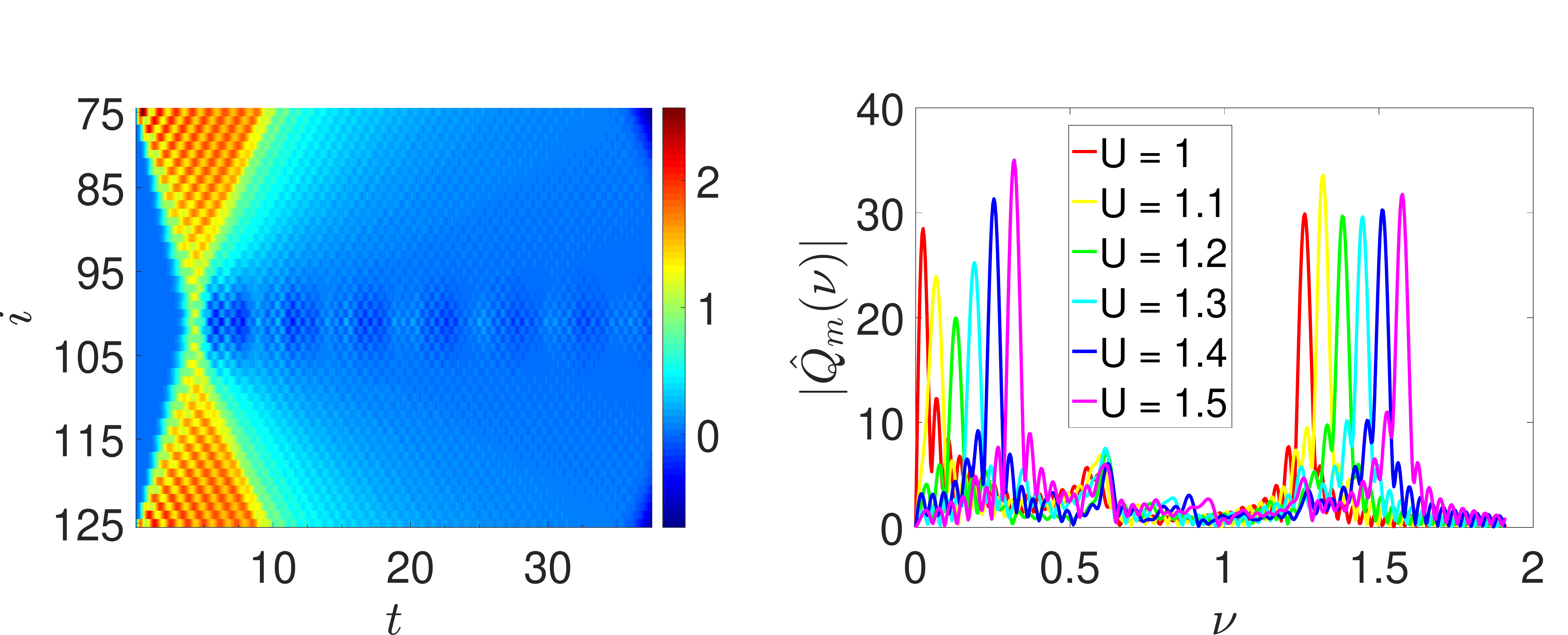}
\caption{Left : Current profile for $U = 1.3$. High and low frequency oscillations are well visible. Right : $|\hat{\tilde{Q}}_m(\nu)|$ for points on phase boundary.} 
\label{Fourier_Insulating}
\end{figure}

To further characterize these oscillations, we investigate the midsection current $Q_{m} = Q_{N_B + \frac{N_S}{2}}$ and the Fourier transform of its oscillations around the mean $\hat{Q}_m(\nu)$. For a system of size $N_S = 50$, we place the beginning of the Fourier analysis at $t = 10/J$. We focus on the homogeneous system $U_S = U_B = U$, which includes points from the ballistic phase as well as the diffusive-insulating phase boundary. Figs.\ref{Fourier_Ballistic} and \ref{Fourier_Insulating} present closeups of the currents in the system for the ballistic phase and diffusive-insulating phase boundary on the left. $\hat{Q}_m(\nu)$ is presented on the right.

We distinguish three main oscillations. The higher frequency peak, with a frequency between 1.2 and 1.6, is responsible for the checkerboard pattern visible in both current pictures. The middle peak, with a frequency between 0.3 and 0.6, is best visible in the ballistic phase, where it is responsible for the larger pattern visible in Fig. \ref{Fourier_Ballistic}. The lower peak, with a frequency between 0 and 0.3, only appears on the phase boundary, and is responsible for the pattern in Fig. \ref{Fourier_Insulating}. It is of much higher amplitude than the medium oscillation, and thus overshadows it in this regime, although all three peaks are discernible in the spectrum.

\begin{figure}
\centering
\includegraphics[scale=0.17]{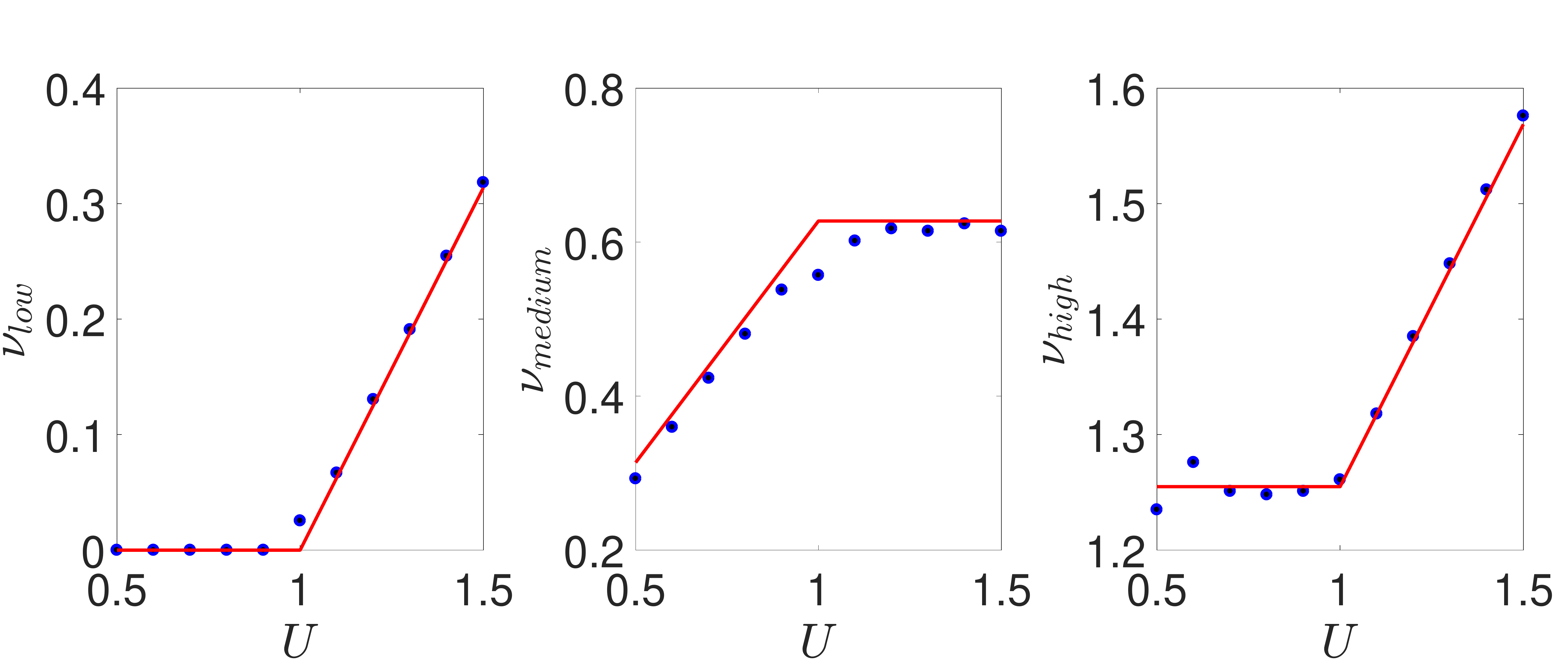}
\caption{One parameter fit of oscillation frequencies. $\alpha = 0.63$} 
\label{Frequency_fit}
\end{figure}

A remarkable feature of these oscillations is their persistence in the long time limit, which is only possible in the diffusive phase due to the very slow relaxation to a stationary state. This feature is of course absent in the phenomenological Master equation description, and is an indicator of the non-Markovian character of the strong coupling regime studied here. As previously mentioned, the low frequency oscillation is unique in that it remains trapped in the middle of the system. The combination of its localization and long-time persistence suggests a rather reminiscent analogy to classical solitons. These non-trivial phenomena highlight the relevance of studying the strong coupling non-Markovian regime from both a theoretical and experimental point of view.

The frequencies of the oscillations are well described by the following expressions :

\begin{align*}
    \nu_{low}&= \alpha \ \max(U-J,0) \\
    \nu_{medium}&= \alpha \ \min(U,J) \\
    \nu_{high}&=\alpha \ (2J + \max(U-J,0))
\end{align*}

Fig. \ref{Frequency_fit} presents the result of fitting the data using a single $\alpha$ for all three frequencies. The qualitative change of frequency scaling is remarkable and coincides with the change from ballistic phase to insulating-diffusive phase boundary.

\section{Conclusion}

We have presented a study of the effects of bath interactions on the transport phases of a non-Markovian boundary-driven spin chain. Behavior for $U_B < 1$ was analogous to previous results for non-interacting baths \cite{Giacommo}. Above $U_B \geq 1$ we have presented evidence for bath interaction induced transitions to a novel diffusive phase which we characterized by power-law finite-size scaling. Most importantly, we have shown that this diffusive phase has a distinctive long-time-algebraic decay of the current $Q\propto t^{-\alpha}$. Along $U_B = U_S$, ballistic and diffusive behavior was found, analogous to results from the single-junction case \cite{SingleJunctionDiffusiveBallistic}. In the homogeneous system, we have characterized the oscillations arising from the non-Markovian aspect of the strong coupling. Their amplitude was found to not decay at long times, and spatial localization of one of the oscillations was observed. The scaling of oscillation frequencies was found to sharply change at the Heisenberg point, coinciding with transition from ballistic phase to insulating-diffusive boundary. These findings attest to the relevance of considering non-Markovian coupling that goes beyond the local-phenomenological master-equation treatment. As a future perspective, it would interesting to further explore the quasi-particle picture to better understand the interference of the light-cones and the nature of the trapped oscillations we have observed. One possible route is the Bethe-based Hydrodynamical approach recently introduced and applied to integrable regimes \cite{Hydrodynamics,Final1,Final2,Final3,Final4}.

\bibliography{references}

\begin{thebibliography}{33}%
\makeatletter
\providecommand \@ifxundefined [1]{%
 \@ifx{#1\undefined}
}%
\providecommand \@ifnum [1]{%
 \ifnum #1\expandafter \@firstoftwo
 \else \expandafter \@secondoftwo
 \fi
}%
\providecommand \@ifx [1]{%
 \ifx #1\expandafter \@firstoftwo
 \else \expandafter \@secondoftwo
 \fi
}%
\providecommand \natexlab [1]{#1}%
\providecommand \enquote  [1]{``#1''}%
\providecommand \bibnamefont  [1]{#1}%
\providecommand \bibfnamefont [1]{#1}%
\providecommand \citenamefont [1]{#1}%
\providecommand \href@noop [0]{\@secondoftwo}%
\providecommand \href [0]{\begingroup \@sanitize@url \@href}%
\providecommand \@href[1]{\@@startlink{#1}\@@href}%
\providecommand \@@href[1]{\endgroup#1\@@endlink}%
\providecommand \@sanitize@url [0]{\catcode `\\12\catcode `\$12\catcode
  `\&12\catcode `\#12\catcode `\^12\catcode `\_12\catcode `\%12\relax}%
\providecommand \@@startlink[1]{}%
\providecommand \@@endlink[0]{}%
\providecommand \url  [0]{\begingroup\@sanitize@url \@url }%
\providecommand \@url [1]{\endgroup\@href {#1}{\urlprefix }}%
\providecommand \urlprefix  [0]{URL }%
\providecommand \Eprint [0]{\href }%
\providecommand \doibase [0]{http://dx.doi.org/}%
\providecommand \selectlanguage [0]{\@gobble}%
\providecommand \bibinfo  [0]{\@secondoftwo}%
\providecommand \bibfield  [0]{\@secondoftwo}%
\providecommand \translation [1]{[#1]}%
\providecommand \BibitemOpen [0]{}%
\providecommand \bibitemStop [0]{}%
\providecommand \bibitemNoStop [0]{.\EOS\space}%
\providecommand \EOS [0]{\spacefactor3000\relax}%
\providecommand \BibitemShut  [1]{\csname bibitem#1\endcsname}%
\let\auto@bib@innerbib\@empty
%</preamble>
\bibitem [{\citenamefont {Calabrese}\ \emph {et~al.}(2016)\citenamefont
  {Calabrese}, \citenamefont {Essler},\ and\ \citenamefont
  {Mussardo}}]{SpecialIssue}%
  \BibitemOpen
  \bibfield  {author} {\bibinfo {author} {\bibfnamefont {P.}~\bibnamefont
  {Calabrese}}, \bibinfo {author} {\bibfnamefont {F.~H.~L.}\ \bibnamefont
  {Essler}}, \ and\ \bibinfo {author} {\bibfnamefont {G.}~\bibnamefont
  {Mussardo}},\ }\href {http://stacks.iop.org/1742-5468/2016/i=6/a=064001}
  {\bibfield  {journal} {\bibinfo  {journal} {Journal of Statistical Mechanics:
  Theory and Experiment}\ }\textbf {\bibinfo {volume} {2016}},\ \bibinfo
  {pages} {064001} (\bibinfo {year} {2016})}\BibitemShut {NoStop}%
\bibitem [{\citenamefont {Polkovnikov}\ \emph {et~al.}(2011)\citenamefont
  {Polkovnikov}, \citenamefont {Sengupta}, \citenamefont {Silva},\ and\
  \citenamefont {Vengalattore}}]{QuenchColloquim}%
  \BibitemOpen
  \bibfield  {author} {\bibinfo {author} {\bibfnamefont {A.}~\bibnamefont
  {Polkovnikov}}, \bibinfo {author} {\bibfnamefont {K.}~\bibnamefont
  {Sengupta}}, \bibinfo {author} {\bibfnamefont {A.}~\bibnamefont {Silva}}, \
  and\ \bibinfo {author} {\bibfnamefont {M.}~\bibnamefont {Vengalattore}},\
  }\href {\doibase 10.1103/RevModPhys.83.863} {\bibfield  {journal} {\bibinfo
  {journal} {Rev. Mod. Phys.}\ }\textbf {\bibinfo {volume} {83}},\ \bibinfo
  {pages} {863} (\bibinfo {year} {2011})}\BibitemShut {NoStop}%
\bibitem [{\citenamefont {Ketterle}\ \emph {et~al.}(1993)\citenamefont
  {Ketterle}, \citenamefont {Davis}, \citenamefont {Joffe}, \citenamefont
  {Martin},\ and\ \citenamefont {Pritchard}}]{Cold1}%
  \BibitemOpen
  \bibfield  {author} {\bibinfo {author} {\bibfnamefont {W.}~\bibnamefont
  {Ketterle}}, \bibinfo {author} {\bibfnamefont {K.~B.}\ \bibnamefont {Davis}},
  \bibinfo {author} {\bibfnamefont {M.~A.}\ \bibnamefont {Joffe}}, \bibinfo
  {author} {\bibfnamefont {A.}~\bibnamefont {Martin}}, \ and\ \bibinfo {author}
  {\bibfnamefont {D.~E.}\ \bibnamefont {Pritchard}},\ }\href {\doibase
  10.1103/PhysRevLett.70.2253} {\bibfield  {journal} {\bibinfo  {journal}
  {Phys. Rev. Lett.}\ }\textbf {\bibinfo {volume} {70}},\ \bibinfo {pages}
  {2253} (\bibinfo {year} {1993})}\BibitemShut {NoStop}%
\bibitem [{\citenamefont {Greiner}\ \emph {et~al.}(2002)\citenamefont
  {Greiner}, \citenamefont {Mandel}, \citenamefont {Esslinger}, \citenamefont
  {Hansch},\ and\ \citenamefont {Bloch}}]{Cold2}%
  \BibitemOpen
  \bibfield  {author} {\bibinfo {author} {\bibfnamefont {M.}~\bibnamefont
  {Greiner}}, \bibinfo {author} {\bibfnamefont {O.}~\bibnamefont {Mandel}},
  \bibinfo {author} {\bibfnamefont {T.}~\bibnamefont {Esslinger}}, \bibinfo
  {author} {\bibfnamefont {T.~W.}\ \bibnamefont {Hansch}}, \ and\ \bibinfo
  {author} {\bibfnamefont {I.}~\bibnamefont {Bloch}},\ }\href {\doibase
  10.1038/415039a} {\bibfield  {journal} {\bibinfo  {journal} {Nature}\
  }\textbf {\bibinfo {volume} {415}},\ \bibinfo {pages} {39} (\bibinfo {year}
  {2002})}\BibitemShut {NoStop}%
\bibitem [{\citenamefont {Kinoshita}\ \emph {et~al.}(2006)\citenamefont
  {Kinoshita}, \citenamefont {Wenger},\ and\ \citenamefont {Weiss}}]{Cold3}%
  \BibitemOpen
  \bibfield  {author} {\bibinfo {author} {\bibfnamefont {T.}~\bibnamefont
  {Kinoshita}}, \bibinfo {author} {\bibfnamefont {T.}~\bibnamefont {Wenger}}, \
  and\ \bibinfo {author} {\bibfnamefont {D.~S.}\ \bibnamefont {Weiss}},\ }\href
  {\doibase 10.1038/nature04693} {\bibfield  {journal} {\bibinfo  {journal}
  {Nature}\ }\textbf {\bibinfo {volume} {440}},\ \bibinfo {pages} {900}
  (\bibinfo {year} {2006})}\BibitemShut {NoStop}%
\bibitem [{\citenamefont {Cheneau}\ \emph {et~al.}(2012)\citenamefont
  {Cheneau}, \citenamefont {Barmettler}, \citenamefont {Poletti}, \citenamefont
  {Endres}, \citenamefont {Schausz}, \citenamefont {Fukuhara}, \citenamefont
  {Gross}, \citenamefont {Bloch}, \citenamefont {Kollath},\ and\ \citenamefont
  {Kuhr}}]{Cold4}%
  \BibitemOpen
  \bibfield  {author} {\bibinfo {author} {\bibfnamefont {M.}~\bibnamefont
  {Cheneau}}, \bibinfo {author} {\bibfnamefont {P.}~\bibnamefont {Barmettler}},
  \bibinfo {author} {\bibfnamefont {D.}~\bibnamefont {Poletti}}, \bibinfo
  {author} {\bibfnamefont {M.}~\bibnamefont {Endres}}, \bibinfo {author}
  {\bibfnamefont {P.}~\bibnamefont {Schausz}}, \bibinfo {author} {\bibfnamefont
  {T.}~\bibnamefont {Fukuhara}}, \bibinfo {author} {\bibfnamefont
  {C.}~\bibnamefont {Gross}}, \bibinfo {author} {\bibfnamefont
  {I.}~\bibnamefont {Bloch}}, \bibinfo {author} {\bibfnamefont
  {C.}~\bibnamefont {Kollath}}, \ and\ \bibinfo {author} {\bibfnamefont
  {S.}~\bibnamefont {Kuhr}},\ }\href {\doibase 10.1038/nature10748} {\bibfield
  {journal} {\bibinfo  {journal} {Nature}\ }\textbf {\bibinfo {volume} {481}},\
  \bibinfo {pages} {484} (\bibinfo {year} {2012})}\BibitemShut {NoStop}%
\bibitem [{\citenamefont {Smith}\ \emph {et~al.}(2016)\citenamefont {Smith},
  \citenamefont {Lee}, \citenamefont {Richerme}, \citenamefont {Neyenhuis},
  \citenamefont {Hess}, \citenamefont {Hauke}, \citenamefont {Heyl},
  \citenamefont {Huse},\ and\ \citenamefont {Monroe}}]{Cold5}%
  \BibitemOpen
  \bibfield  {author} {\bibinfo {author} {\bibfnamefont {J.}~\bibnamefont
  {Smith}}, \bibinfo {author} {\bibfnamefont {A.}~\bibnamefont {Lee}}, \bibinfo
  {author} {\bibfnamefont {P.}~\bibnamefont {Richerme}}, \bibinfo {author}
  {\bibfnamefont {B.}~\bibnamefont {Neyenhuis}}, \bibinfo {author}
  {\bibfnamefont {P.~W.}\ \bibnamefont {Hess}}, \bibinfo {author}
  {\bibfnamefont {P.}~\bibnamefont {Hauke}}, \bibinfo {author} {\bibfnamefont
  {M.}~\bibnamefont {Heyl}}, \bibinfo {author} {\bibfnamefont {D.~A.}\
  \bibnamefont {Huse}}, \ and\ \bibinfo {author} {\bibfnamefont
  {C.}~\bibnamefont {Monroe}},\ }\href {http://dx.doi.org/10.1038/nphys3783}
  {\bibfield  {journal} {\bibinfo  {journal} {Nat Phys}\ }\textbf {\bibinfo
  {volume} {12}},\ \bibinfo {pages} {907} (\bibinfo {year} {2016})},\ \bibinfo
  {note} {letter}\BibitemShut {NoStop}%
\bibitem [{\citenamefont {Schweizer}\ \emph {et~al.}(2016)\citenamefont
  {Schweizer}, \citenamefont {Lohse}, \citenamefont {Citro},\ and\
  \citenamefont {Bloch}}]{Cold6}%
  \BibitemOpen
  \bibfield  {author} {\bibinfo {author} {\bibfnamefont {C.}~\bibnamefont
  {Schweizer}}, \bibinfo {author} {\bibfnamefont {M.}~\bibnamefont {Lohse}},
  \bibinfo {author} {\bibfnamefont {R.}~\bibnamefont {Citro}}, \ and\ \bibinfo
  {author} {\bibfnamefont {I.}~\bibnamefont {Bloch}},\ }\href {\doibase
  10.1103/PhysRevLett.117.170405} {\bibfield  {journal} {\bibinfo  {journal}
  {Phys. Rev. Lett.}\ }\textbf {\bibinfo {volume} {117}},\ \bibinfo {pages}
  {170405} (\bibinfo {year} {2016})}\BibitemShut {NoStop}%
\bibitem [{\citenamefont {Sachdev}(2011)}]{QuantumPhase}%
  \BibitemOpen
  \bibfield  {author} {\bibinfo {author} {\bibfnamefont {S.}~\bibnamefont
  {Sachdev}},\ }\href@noop {} {\emph {\bibinfo {title} {Quantum phase
  transitions}}},\ \bibinfo {edition} {second ed.}\ ed.\ (\bibinfo  {publisher}
  {Cambridge University Press},\ \bibinfo {address} {Cambridge},\ \bibinfo
  {year} {2011})\BibitemShut {NoStop}%
\bibitem [{\citenamefont {Haken}(1983)}]{Synergetics}%
  \BibitemOpen
  \bibfield  {author} {\bibinfo {author} {\bibfnamefont {H.}~\bibnamefont
  {Haken}},\ }\href@noop {} {\emph {\bibinfo {title} {{Synergetics, An
  Introduction}}}},\ \bibinfo {edition} {3rd}\ ed.\ (\bibinfo  {publisher}
  {Springer},\ \bibinfo {address} {Berlin},\ \bibinfo {year}
  {1983})\BibitemShut {NoStop}%
\bibitem [{\citenamefont {Muller}\ \emph {et~al.}(1981)\citenamefont {Muller},
  \citenamefont {Thomas}, \citenamefont {Puga},\ and\ \citenamefont
  {Beck}}]{XXZ1}%
  \BibitemOpen
  \bibfield  {author} {\bibinfo {author} {\bibfnamefont {G.}~\bibnamefont
  {Muller}}, \bibinfo {author} {\bibfnamefont {H.}~\bibnamefont {Thomas}},
  \bibinfo {author} {\bibfnamefont {M.~W.}\ \bibnamefont {Puga}}, \ and\
  \bibinfo {author} {\bibfnamefont {H.}~\bibnamefont {Beck}},\ }\href
  {http://stacks.iop.org/0022-3719/14/i=23/a=017} {\bibfield  {journal}
  {\bibinfo  {journal} {Journal of Physics C: Solid State Physics}\ }\textbf
  {\bibinfo {volume} {14}},\ \bibinfo {pages} {3399} (\bibinfo {year}
  {1981})}\BibitemShut {NoStop}%
\bibitem [{\citenamefont {Sologubenko}\ \emph {et~al.}(2007)\citenamefont
  {Sologubenko}, \citenamefont {Lorenz}, \citenamefont {Ott},\ and\
  \citenamefont {Freimuth}}]{XXZ2}%
  \BibitemOpen
  \bibfield  {author} {\bibinfo {author} {\bibfnamefont {A.~V.}\ \bibnamefont
  {Sologubenko}}, \bibinfo {author} {\bibfnamefont {T.}~\bibnamefont {Lorenz}},
  \bibinfo {author} {\bibfnamefont {H.~R.}\ \bibnamefont {Ott}}, \ and\
  \bibinfo {author} {\bibfnamefont {A.}~\bibnamefont {Freimuth}},\ }\href
  {\doibase 10.1007/s10909-007-9317-x} {\bibfield  {journal} {\bibinfo
  {journal} {Journal of Low Temperature Physics}\ }\textbf {\bibinfo {volume}
  {147}},\ \bibinfo {pages} {387} (\bibinfo {year} {2007})}\BibitemShut
  {NoStop}%
\bibitem [{\citenamefont {Hirobe}\ \emph {et~al.}(2017)\citenamefont {Hirobe},
  \citenamefont {Sato}, \citenamefont {Kawamata}, \citenamefont {Shiomi},
  \citenamefont {Uchida}, \citenamefont {Iguchi}, \citenamefont {Koike},
  \citenamefont {Maekawa},\ and\ \citenamefont {Saitoh}}]{XXZ3}%
  \BibitemOpen
  \bibfield  {author} {\bibinfo {author} {\bibfnamefont {D.}~\bibnamefont
  {Hirobe}}, \bibinfo {author} {\bibfnamefont {M.}~\bibnamefont {Sato}},
  \bibinfo {author} {\bibfnamefont {T.}~\bibnamefont {Kawamata}}, \bibinfo
  {author} {\bibfnamefont {Y.}~\bibnamefont {Shiomi}}, \bibinfo {author}
  {\bibfnamefont {K.-i.}\ \bibnamefont {Uchida}}, \bibinfo {author}
  {\bibfnamefont {R.}~\bibnamefont {Iguchi}}, \bibinfo {author} {\bibfnamefont
  {Y.}~\bibnamefont {Koike}}, \bibinfo {author} {\bibfnamefont
  {S.}~\bibnamefont {Maekawa}}, \ and\ \bibinfo {author} {\bibfnamefont
  {E.}~\bibnamefont {Saitoh}},\ }\href {http://dx.doi.org/10.1038/nphys3895}
  {\bibfield  {journal} {\bibinfo  {journal} {Nat Phys}\ }\textbf {\bibinfo
  {volume} {13}},\ \bibinfo {pages} {30} (\bibinfo {year} {2017})},\ \bibinfo
  {note} {letter}\BibitemShut {NoStop}%
\bibitem [{\citenamefont {\ifmmode \check{Z}\else
  \v{Z}\fi{}nidari\ifmmode~\check{c}\else
  \v{c}\fi{}}(2011)}]{ZnidaricLinearResponse}%
  \BibitemOpen
  \bibfield  {author} {\bibinfo {author} {\bibfnamefont {M.}~\bibnamefont
  {\ifmmode \check{Z}\else \v{Z}\fi{}nidari\ifmmode~\check{c}\else
  \v{c}\fi{}}},\ }\href {\doibase 10.1103/PhysRevLett.106.220601} {\bibfield
  {journal} {\bibinfo  {journal} {Phys. Rev. Lett.}\ }\textbf {\bibinfo
  {volume} {106}},\ \bibinfo {pages} {220601} (\bibinfo {year}
  {2011})}\BibitemShut {NoStop}%
\bibitem [{\citenamefont {Prosen}(2011)}]{ProsenExactNess}%
  \BibitemOpen
  \bibfield  {author} {\bibinfo {author} {\bibfnamefont {T.~c.~v.}\
  \bibnamefont {Prosen}},\ }\href {\doibase 10.1103/PhysRevLett.107.137201}
  {\bibfield  {journal} {\bibinfo  {journal} {Phys. Rev. Lett.}\ }\textbf
  {\bibinfo {volume} {107}},\ \bibinfo {pages} {137201} (\bibinfo {year}
  {2011})}\BibitemShut {NoStop}%
\bibitem [{\citenamefont {Breuer}\ and\ \citenamefont
  {Petruccione}(2002)}]{BreuerPetruccione}%
  \BibitemOpen
  \bibfield  {author} {\bibinfo {author} {\bibfnamefont {H.}~\bibnamefont
  {Breuer}}\ and\ \bibinfo {author} {\bibfnamefont {F.}~\bibnamefont
  {Petruccione}},\ }\href {https://books.google.ch/books?id=w2UOnwEACAAJ}
  {\emph {\bibinfo {title} {The Theory of Open Quantum Systems}}}\ (\bibinfo
  {publisher} {Oxford University Press},\ \bibinfo {year} {2002})\BibitemShut
  {NoStop}%
\bibitem [{\citenamefont {{Rivas}}\ and\ \citenamefont
  {{Martin-Delgado}}(2016)}]{ChiralCurrents}%
  \BibitemOpen
  \bibfield  {author} {\bibinfo {author} {\bibfnamefont {{\'A}.}~\bibnamefont
  {{Rivas}}}\ and\ \bibinfo {author} {\bibfnamefont {M.~A.}\ \bibnamefont
  {{Martin-Delgado}}},\ }\href@noop {} {\bibfield  {journal} {\bibinfo
  {journal} {ArXiv e-prints}\ } (\bibinfo {year} {2016})},\ \Eprint
  {http://arxiv.org/abs/1606.07651} {arXiv:1606.07651 [cond-mat.mes-hall]}
  \BibitemShut {NoStop}%
\bibitem [{\citenamefont {Ljubotina}\ \emph {et~al.}(2017)\citenamefont
  {Ljubotina}, \citenamefont {Znidaric},\ and\ \citenamefont
  {Prosen}}]{SolvenianSingleJunction}%
  \BibitemOpen
  \bibfield  {author} {\bibinfo {author} {\bibfnamefont {M.}~\bibnamefont
  {Ljubotina}}, \bibinfo {author} {\bibfnamefont {M.}~\bibnamefont {Znidaric}},
  \ and\ \bibinfo {author} {\bibfnamefont {T.}~\bibnamefont {Prosen}},\ }\href
  {http://dx.doi.org/10.1038/ncomms16117} {\bibfield  {journal} {\bibinfo
  {journal} {Nature Communications}\ }\textbf {\bibinfo {volume} {8}} (\bibinfo
  {year} {2017})}\BibitemShut {NoStop}%
\bibitem [{\citenamefont {Bonnes}\ \emph {et~al.}(2014)\citenamefont {Bonnes},
  \citenamefont {Essler},\ and\ \citenamefont {L\"auchli}}]{LightCone}%
  \BibitemOpen
  \bibfield  {author} {\bibinfo {author} {\bibfnamefont {L.}~\bibnamefont
  {Bonnes}}, \bibinfo {author} {\bibfnamefont {F.~H.~L.}\ \bibnamefont
  {Essler}}, \ and\ \bibinfo {author} {\bibfnamefont {A.~M.}\ \bibnamefont
  {L\"auchli}},\ }\href {\doibase 10.1103/PhysRevLett.113.187203} {\bibfield
  {journal} {\bibinfo  {journal} {Phys. Rev. Lett.}\ }\textbf {\bibinfo
  {volume} {113}},\ \bibinfo {pages} {187203} (\bibinfo {year}
  {2014})}\BibitemShut {NoStop}%
\bibitem [{\citenamefont {Alba}\ and\ \citenamefont
  {Heidrich-Meisner}(2014)}]{EntanglementSpreading}%
  \BibitemOpen
  \bibfield  {author} {\bibinfo {author} {\bibfnamefont {V.}~\bibnamefont
  {Alba}}\ and\ \bibinfo {author} {\bibfnamefont {F.}~\bibnamefont
  {Heidrich-Meisner}},\ }\href {\doibase 10.1103/PhysRevB.90.075144} {\bibfield
   {journal} {\bibinfo  {journal} {Phys. Rev. B}\ }\textbf {\bibinfo {volume}
  {90}},\ \bibinfo {pages} {075144} (\bibinfo {year} {2014})}\BibitemShut
  {NoStop}%
\bibitem [{\citenamefont {De~Luca}\ \emph {et~al.}(2014)\citenamefont
  {De~Luca}, \citenamefont {Viti}, \citenamefont {Mazza},\ and\ \citenamefont
  {Rossini}}]{Thermal1}%
  \BibitemOpen
  \bibfield  {author} {\bibinfo {author} {\bibfnamefont {A.}~\bibnamefont
  {De~Luca}}, \bibinfo {author} {\bibfnamefont {J.}~\bibnamefont {Viti}},
  \bibinfo {author} {\bibfnamefont {L.}~\bibnamefont {Mazza}}, \ and\ \bibinfo
  {author} {\bibfnamefont {D.}~\bibnamefont {Rossini}},\ }\href {\doibase
  10.1103/PhysRevB.90.161101} {\bibfield  {journal} {\bibinfo  {journal} {Phys.
  Rev. B}\ }\textbf {\bibinfo {volume} {90}},\ \bibinfo {pages} {161101}
  (\bibinfo {year} {2014})}\BibitemShut {NoStop}%
\bibitem [{\citenamefont {Collura}\ and\ \citenamefont
  {Martelloni}(2014)}]{Thermal2}%
  \BibitemOpen
  \bibfield  {author} {\bibinfo {author} {\bibfnamefont {M.}~\bibnamefont
  {Collura}}\ and\ \bibinfo {author} {\bibfnamefont {G.}~\bibnamefont
  {Martelloni}},\ }\href {http://stacks.iop.org/1742-5468/2014/i=8/a=P08006}
  {\bibfield  {journal} {\bibinfo  {journal} {Journal of Statistical Mechanics:
  Theory and Experiment}\ }\textbf {\bibinfo {volume} {2014}},\ \bibinfo
  {pages} {P08006} (\bibinfo {year} {2014})}\BibitemShut {NoStop}%
\bibitem [{\citenamefont {Biella}\ \emph {et~al.}(2016)\citenamefont {Biella},
  \citenamefont {De~Luca}, \citenamefont {Viti}, \citenamefont {Rossini},
  \citenamefont {Mazza},\ and\ \citenamefont {Fazio}}]{Thermal3}%
  \BibitemOpen
  \bibfield  {author} {\bibinfo {author} {\bibfnamefont {A.}~\bibnamefont
  {Biella}}, \bibinfo {author} {\bibfnamefont {A.}~\bibnamefont {De~Luca}},
  \bibinfo {author} {\bibfnamefont {J.}~\bibnamefont {Viti}}, \bibinfo {author}
  {\bibfnamefont {D.}~\bibnamefont {Rossini}}, \bibinfo {author} {\bibfnamefont
  {L.}~\bibnamefont {Mazza}}, \ and\ \bibinfo {author} {\bibfnamefont
  {R.}~\bibnamefont {Fazio}},\ }\href {\doibase 10.1103/PhysRevB.93.205121}
  {\bibfield  {journal} {\bibinfo  {journal} {Phys. Rev. B}\ }\textbf {\bibinfo
  {volume} {93}},\ \bibinfo {pages} {205121} (\bibinfo {year}
  {2016})}\BibitemShut {NoStop}%
\bibitem [{\citenamefont {Karrasch}\ \emph {et~al.}(2013)\citenamefont
  {Karrasch}, \citenamefont {Ilan},\ and\ \citenamefont {Moore}}]{Thermal4}%
  \BibitemOpen
  \bibfield  {author} {\bibinfo {author} {\bibfnamefont {C.}~\bibnamefont
  {Karrasch}}, \bibinfo {author} {\bibfnamefont {R.}~\bibnamefont {Ilan}}, \
  and\ \bibinfo {author} {\bibfnamefont {J.~E.}\ \bibnamefont {Moore}},\ }\href
  {\doibase 10.1103/PhysRevB.88.195129} {\bibfield  {journal} {\bibinfo
  {journal} {Phys. Rev. B}\ }\textbf {\bibinfo {volume} {88}},\ \bibinfo
  {pages} {195129} (\bibinfo {year} {2013})}\BibitemShut {NoStop}%
\bibitem [{\citenamefont {Collura}\ and\ \citenamefont
  {Karevski}(2014)}]{Thermal5}%
  \BibitemOpen
  \bibfield  {author} {\bibinfo {author} {\bibfnamefont {M.}~\bibnamefont
  {Collura}}\ and\ \bibinfo {author} {\bibfnamefont {D.}~\bibnamefont
  {Karevski}},\ }\href {\doibase 10.1103/PhysRevB.89.214308} {\bibfield
  {journal} {\bibinfo  {journal} {Phys. Rev. B}\ }\textbf {\bibinfo {volume}
  {89}},\ \bibinfo {pages} {214308} (\bibinfo {year} {2014})}\BibitemShut
  {NoStop}%
\bibitem [{\citenamefont {Castro-Alvaredo}\ \emph {et~al.}(2016)\citenamefont
  {Castro-Alvaredo}, \citenamefont {Doyon},\ and\ \citenamefont
  {Yoshimura}}]{Hydrodynamics}%
  \BibitemOpen
  \bibfield  {author} {\bibinfo {author} {\bibfnamefont {O.~A.}\ \bibnamefont
  {Castro-Alvaredo}}, \bibinfo {author} {\bibfnamefont {B.}~\bibnamefont
  {Doyon}}, \ and\ \bibinfo {author} {\bibfnamefont {T.}~\bibnamefont
  {Yoshimura}},\ }\href {\doibase 10.1103/PhysRevX.6.041065} {\bibfield
  {journal} {\bibinfo  {journal} {Phys. Rev. X}\ }\textbf {\bibinfo {volume}
  {6}},\ \bibinfo {pages} {041065} (\bibinfo {year} {2016})}\BibitemShut
  {NoStop}%
\bibitem [{\citenamefont {Mascarenhas}\ \emph {et~al.}(2017)\citenamefont
  {Mascarenhas}, \citenamefont {Giudice},\ and\ \citenamefont
  {Savona}}]{Giacommo}%
  \BibitemOpen
  \bibfield  {author} {\bibinfo {author} {\bibfnamefont {E.}~\bibnamefont
  {Mascarenhas}}, \bibinfo {author} {\bibfnamefont {G.}~\bibnamefont
  {Giudice}}, \ and\ \bibinfo {author} {\bibfnamefont {V.}~\bibnamefont
  {Savona}},\ }\href {https://arxiv.org/pdf/1703.02934.pdf} {\bibfield
  {journal} {\bibinfo  {journal} {arXiv preprint arXiv:1703.02934}\ } (\bibinfo
  {year} {2017})}\BibitemShut {NoStop}%
\bibitem [{\citenamefont {\ifmmode \check{Z}\else
  \v{Z}\fi{}nidari\ifmmode~\check{c}\else \v{c}\fi{}}\ \emph
  {et~al.}(2016)\citenamefont {\ifmmode \check{Z}\else
  \v{Z}\fi{}nidari\ifmmode~\check{c}\else \v{c}\fi{}}, \citenamefont
  {Scardicchio},\ and\ \citenamefont {Varma}}]{PhysRevLett.117.040601}%
  \BibitemOpen
  \bibfield  {author} {\bibinfo {author} {\bibfnamefont {M.}~\bibnamefont
  {\ifmmode \check{Z}\else \v{Z}\fi{}nidari\ifmmode~\check{c}\else
  \v{c}\fi{}}}, \bibinfo {author} {\bibfnamefont {A.}~\bibnamefont
  {Scardicchio}}, \ and\ \bibinfo {author} {\bibfnamefont {V.~K.}\ \bibnamefont
  {Varma}},\ }\href {\doibase 10.1103/PhysRevLett.117.040601} {\bibfield
  {journal} {\bibinfo  {journal} {Phys. Rev. Lett.}\ }\textbf {\bibinfo
  {volume} {117}},\ \bibinfo {pages} {040601} (\bibinfo {year}
  {2016})}\BibitemShut {NoStop}%
\bibitem [{\citenamefont {Karrasch}\ \emph {et~al.}(2014)\citenamefont
  {Karrasch}, \citenamefont {Moore},\ and\ \citenamefont
  {Heidrich-Meisner}}]{SingleJunctionDiffusiveBallistic}%
  \BibitemOpen
  \bibfield  {author} {\bibinfo {author} {\bibfnamefont {C.}~\bibnamefont
  {Karrasch}}, \bibinfo {author} {\bibfnamefont {J.~E.}\ \bibnamefont {Moore}},
  \ and\ \bibinfo {author} {\bibfnamefont {F.}~\bibnamefont
  {Heidrich-Meisner}},\ }\href {\doibase 10.1103/PhysRevB.89.075139} {\bibfield
   {journal} {\bibinfo  {journal} {Phys. Rev. B}\ }\textbf {\bibinfo {volume}
  {89}},\ \bibinfo {pages} {075139} (\bibinfo {year} {2014})}\BibitemShut
  {NoStop}%
\bibitem [{\citenamefont {Alba}(2017)}]{Final1}%
  \BibitemOpen
  \bibfield  {author} {\bibinfo {author} {\bibfnamefont {V.}~\bibnamefont
  {Alba}},\ }\href {https://arxiv.org/abs/1706.00020.pdf} {\bibfield  {journal}
  {\bibinfo  {journal} {arXiv preprint arXiv:1706.00020}\ } (\bibinfo {year}
  {2017})}\BibitemShut {NoStop}%
\bibitem [{\citenamefont {Bertini}\ \emph {et~al.}(2016)\citenamefont
  {Bertini}, \citenamefont {Collura}, \citenamefont {De~Nardis},\ and\
  \citenamefont {Fagotti}}]{Final2}%
  \BibitemOpen
  \bibfield  {author} {\bibinfo {author} {\bibfnamefont {B.}~\bibnamefont
  {Bertini}}, \bibinfo {author} {\bibfnamefont {M.}~\bibnamefont {Collura}},
  \bibinfo {author} {\bibfnamefont {J.}~\bibnamefont {De~Nardis}}, \ and\
  \bibinfo {author} {\bibfnamefont {M.}~\bibnamefont {Fagotti}},\ }\href
  {\doibase 10.1103/PhysRevLett.117.207201} {\bibfield  {journal} {\bibinfo
  {journal} {Phys. Rev. Lett.}\ }\textbf {\bibinfo {volume} {117}},\ \bibinfo
  {pages} {207201} (\bibinfo {year} {2016})}\BibitemShut {NoStop}%
\bibitem [{\citenamefont {Doyon}\ and\ \citenamefont
  {Yoshimura}(2016)}]{Final3}%
  \BibitemOpen
  \bibfield  {author} {\bibinfo {author} {\bibfnamefont {B.}~\bibnamefont
  {Doyon}}\ and\ \bibinfo {author} {\bibfnamefont {T.}~\bibnamefont
  {Yoshimura}},\ }\href {https://arxiv.org/abs/1611.08225.pdf} {\bibfield
  {journal} {\bibinfo  {journal} {arXiv preprint arXiv:1611.08225}\ } (\bibinfo
  {year} {2016})}\BibitemShut {NoStop}%
\bibitem [{\citenamefont {De~Luca}\ \emph {et~al.}(2017)\citenamefont
  {De~Luca}, \citenamefont {Collura},\ and\ \citenamefont
  {De~Nardis}}]{Final4}%
  \BibitemOpen
  \bibfield  {author} {\bibinfo {author} {\bibfnamefont {A.}~\bibnamefont
  {De~Luca}}, \bibinfo {author} {\bibfnamefont {M.}~\bibnamefont {Collura}}, \
  and\ \bibinfo {author} {\bibfnamefont {J.}~\bibnamefont {De~Nardis}},\ }\href
  {\doibase 10.1103/PhysRevB.96.020403} {\bibfield  {journal} {\bibinfo
  {journal} {Phys. Rev. B}\ }\textbf {\bibinfo {volume} {96}},\ \bibinfo
  {pages} {020403} (\bibinfo {year} {2017})}\BibitemShut {NoStop}%
\end{thebibliography}%
\end{document}